\documentclass[prd,twocolumn,showpacs,floatfix,amsmath,nofootinbib,amssymb,floatfix]{revtex4}
\usepackage{graphicx,color,dcolumn,booktabs,bm}
\usepackage{longtable,lscape}
\usepackage{txfonts}
\usepackage{overpic}
\usepackage{amssymb}
\usepackage{indentfirst}
\usepackage{feynmf}   
\usepackage{slashed}  
\usepackage{cases}
\usepackage{color}
\usepackage{multirow}
\usepackage{epstopdf}
\usepackage{longtable}
\usepackage{graphicx,color,dcolumn,booktabs,bm}
\usepackage[colorlinks,
            citecolor=blue,
            anchorcolor=red,
            menucolor=red,
            linkcolor=red,
            filecolor=red,
            runcolor=red,
            urlcolor=blue,
            frenchlinks=red]{hyperref}

\graphicspath{{Figures/}} %

\allowdisplaybreaks

\begin{document}

\title{Possible open charm molecular pentaquarks from $\Lambda_cK^{(*)}/\Sigma_cK^{(*)}$ interactions}

\author{Rui Chen$^{1}$}\email{chenrui@hunnu.edu.cn}
\author{Qi Huang$^{2}$\footnote{Corresponding author}}\email{06289@njnu.edu.cn}

\affiliation{
$^1$Key Laboratory of Low-Dimensional Quantum Structures and Quantum Control of Ministry of Education, Department of Physics and Synergetic Innovation Center for Quantum Effects and Applications, Hunan Normal University, Changsha 410081, China\\
$^2$Department of Physics, Nanjing Normal University, Nanjing 210023, China}
\date{\today}

\begin{abstract}
In this work, we adopt the one-boson-exchange model to study the $Y_cK^{(*)} (Y_c=\Lambda_c, \Sigma_c)$ interactions. After considering both of the $S-D$ wave mixing effects and the coupled channel effects, we can predict several possible open-charm molecular pentaquarks, i.e., the single $\Sigma_cK^*$ molecular states with $I(J^P)=1/2(1/2^-)$, $1/2(3/2^-)$ and $3/2(1/2^-)$, the coupled $\Lambda_cK^*/\Sigma_cK^*$ molecular states with $1/2(1/2^-)$ and $1/2(3/2^-)$, and the coupled $\Sigma_cK/\Lambda_cK^*/\Sigma_cK^*$ molecular state with $1/2(1/2^-)$. Meanwhile, we extend our study to the $Y_c\bar{K}^{(*)}$ interactions, our results suggest the $\Sigma_c\bar{K}$ system with $I(J^P)=1/2(1/2^-)$, the $\Sigma_c\bar{K}^*$ systems with $1/2(1/2^-)$, $1/2(3/2^-)$, and $3/2(3/2^-)$, the coupled $\Lambda_c\bar{K}^*/\Sigma_c\bar K^*$ system with $1/2(1/2^-)$, and the $\Sigma_c\bar{K}/\Lambda_c\bar{K}^*/\Sigma_c\bar K^*$ system with $1/2(1/2^-)$ can be the prime molecular candidates.

\end{abstract}


\maketitle

\section{introduction}

In the past decades, the observations of $X/Y/Z/P_c/T_{cc}$ structures have stimulated theorist's extensive interest in
exploring the properties of exotic states. Among the possible configurations, the hadronic molecular state, which is composed of the color-singlet hadrons, plays an important role in explaining the observed exotic structures. The main reason of introducing such a configuration is that many observed $X/Y/Z/P_c/T_{cc}$ structures are near some specific mass thresholds of the hadron pairs, which leads to answers whether these observations can be explained under the framework of the molecular state (one can see Refs. \cite{Chen:2016qju,Liu:2019zoy,Chen:2016spr,Guo:2017jvc,Chen:2022asf} for a detailed review). Thus, carrying out the study of the hadronic molecular state has became an active and important research field in the hadron physics. It is not only helpful to reveal the underlying structures of these near thresholds $X/Y/Z/P_c/T_{cc}$ structures, but also can improve our knowledge of the non-perturbative behavior of the quantum chromodynamics (QCD).

Very recently, the LHCb collaboration continued to report their observations of two open heavy flavor multiquark candidates, $T_{c\bar{s}}^{a0}(2900)$ and $T_{c\bar{s}}^{a++}(2900)$, where the superscript $a$ means that their quantum numbers are both $I(J^P)=1(0^+)$ \cite{LHCb:Tcc,LHCb:Qian}. For the $T_{c\bar{s}}^{a0}(2900)$, the discovered channel is $D_s^+ \pi^-$, the mass and width are $2892 \pm 14 \pm 15$ MeV and $119 \pm 26 \pm 12$ MeV, respectively, while for the $T_{c\bar{s}}^{a++}(2900)$, the discovered channel, the mass, and the width are $D_s^+ \pi^+$, $2921 \pm 17 \pm 19$ MeV and $137 \pm 32 \pm 14$ MeV, respectively. According to their channels, mass positions and quantum numbers, it is easy to guess that the $T_{c\bar{s}}^{a0}(2900)$ and $T_{c\bar{s}}^{a++}(2900)$ belong to the same isovector triplet. Furthermore, the LHCb collaboration also determined their averaged masses and decay widths, which are $2908 \pm 11 \pm 20$ MeV and $136 \pm 23 \pm 11$ MeV, respectively.

Due to the charged property of $T_{c\bar{s}}^{a0(++)}(2900)$, their minimal valance quark components are naturally inferred to be $c\bar{s}q\bar{q}$ ($q=u,~d$). Since they are very close to the $D^*K^*$ mass threshold, it is natural conjecture whether the $T_{c\bar{s}}^{a0(++)}(2900)$ states can be the isovector $D^*K^*$ molecules with $J^P=0^+$. In fact, in our former work~\cite{Chen:2016ypj}, we can not only reproduce the $D_{s0}^\ast(2317)$ and $D_{s1}(2460)$ in the $S-$wave $DK$ and $D^*K$ molecular scenario, but also find the one-boson-exchange (OBE) effective potentials are strong enough to form loosely bound molecular states for the $D^*K^*$ systems with $I(J^P)=0(0^+, 1^+, 2^+)$, and $1(0^+)$. Therefore, the $D^*K^*$ hadronic molecular explanations for the $T_{c\bar{s}}^{a0(++)}(2900)$ states cannot be excluded. In addition, there are other different theoretical explanations to the $T_{c\bar{s}}^{a0(++)}(2900)$ states, like the compact open-charm pentaquark~\cite{Chen:2017rhl,Guo:2021mja,Cheng:2020nho} and the $D^*\rho$ molecule~\cite{Agaev:2022duz}.

Besides the $T_{c\bar{s}}^{a0(++)}(2900)$, another two open-charm states $X_0(2900)$ and $X_1(2900)$, which were observed by the LHCb collaboration in the $D^-K^+$ final states of the $B^+\to D^+D^-K^+$ decay process~\cite{LHCb:2020bls,LHCb:2020pxc}, are also interesting. Their spin-parities $J^P$ are $0^+$ and $1^+$, respectively. Because their mass positions are very close to the $\bar{D}^*K^*$ and $\bar{D}_1K$ mass thresholds, respectively, many theorists propose the $X_0(2900)$ and $X_1(2900)$ states as the hadronic molecular states~\cite{Molina:2020hde,Hu:2020mxp,Liu:2020nil,Kong:2021ohg,Wang:2021lwy,
Xiao:2020ltm,Huang:2020ptc,Chen:2020aos,Agaev:2020nrc,Qi:2021iyv,Chen:2021tad}. At present, the inner structures for the $T_{c\bar{s}}^{a0(++)}(2900)$ and $X_{0,1}(2900)$ are still on discussion~(one can see Refs. \cite{Chen:2022asf}).

As is well known, the light diquark in the heavy baryons $Y_c=(\Lambda_c, \Sigma_c)$ has the same color structure $\bar{3}_c$ with the light anti-quark in the heavy meson $Q\bar q$ \cite{An:2022vtg}. If the $T_{c\bar{s}}^{a0(++)}(2900)$ can be assigned as the loosely bound hadronic molecular states composed by the charmed meson and kaon, it is natural to conjecture whether there exist possible open charm molecular pentaquarks counterpart of the $T_{c\bar{s}}^{a0(++)}(2900)$, which are near the thresholds of the $\Lambda_cK^{(*)}$ and $\Sigma_cK^{(*)}$, respectively. In this work, we search for such open charm molecular partners composed by $\Lambda_cK^{(*)}$ and $\Sigma_cK^{(*)}$, which can not only enrich the family of the exotic states, but also help us to understand the nature of the newly $T_{c\bar{s}}^{a0(++)}(2900)$.

Apart from searching for possible $\Lambda_cK^{(*)}$ and $\Sigma_cK^{(*)}$ molecular states, in this work, we also study the interactions between the $S-$wave charmed baryon $Y_c=(\Lambda_c, \Sigma_c)$ and the anti-strange meson $\bar{K}^{(*)}$ by adopting the OBE model and considering both of the S-D mixing effects and the coupled channel effects. After solving the coupled channel Schr\"{o}dinger equations, we can search for the possible charmed-strange molecular pentaquarks counterpart of the $X_{0,1}(2900)$. Our study will not only provide valuable information to experimental search for exotic open charm hadronic molecular pentaquarks, but also give indirect test of molecular state picture for the $T_{c\bar{s}}^{a0(++)}(2900)$ and $X_{0,1}(2900)$.

This paper is organized as follows. After this introduction, we introduce the relevant effective Lagrangians and the OBE model in Sec.~\ref{sec2}. In Sec.~\ref{sec3}, we present the OBE effective potentials and the corresponding numerical results. The paper ends with a summary in Sec. \ref{sec4}.

\section{Lagrangians and OBE model}\label{sec2}

In this work, we deduce the OBE effective potentials for the $Y_cK^{(*)}$ systems by employing the effective Lagrangian approach at the hadronic level. The relevant Lagrangians describing the interactions between the heavy baryons and light mesons are constructed in terms of the heavy quark limit and chiral symmetry \cite{Liu:2011xc}, i.e.,
\begin{eqnarray}
\mathcal{L}_{\mathcal{B}_{\bar{3}}} &=& l_B\langle\bar{\mathcal{B}}_{\bar{3}}\sigma\mathcal{B}_{\bar{3}}\rangle
          +i\beta_B\langle\bar{\mathcal{B}}_{\bar{3}}v^{\mu}(\mathcal{V}_{\mu}-\rho_{\mu})\mathcal{B}_{\bar{3}}\rangle,\label{lag1}\\
\mathcal{L}_{\mathcal{B}_{6}} &=&  l_S\langle\bar{\mathcal{S}}_{\mu}\sigma\mathcal{S}^{\mu}\rangle
         -\frac{3}{2}g_1\varepsilon^{\mu\nu\lambda\kappa}v_{\kappa}
         \langle\bar{\mathcal{S}}_{\mu}A_{\nu}\mathcal{S}_{\lambda}\rangle\nonumber\\
    &&+i\beta_{S}\langle\bar{\mathcal{S}}_{\mu}v_{\alpha}
    \left(\mathcal{V}_{ab}^{\alpha}-\rho_{ab}^{\alpha}\right) \mathcal{S}^{\mu}\rangle
    +\lambda_S\langle\bar{\mathcal{S}}_{\mu}F^{\mu\nu}(\rho)\mathcal{S}_{\nu}\rangle,~\\
\mathcal{L}_{\mathcal{B}_{\bar{3}}\mathcal{B}_6} &=& ig_4\langle\bar{\mathcal{S}^{\mu}}A_{\mu}\mathcal{B}_{\bar{3}}\rangle
         +i\lambda_I\varepsilon^{\mu\nu\lambda\kappa}v_{\mu}\langle \bar{\mathcal{S}}_{\nu}F_{\lambda\kappa}\mathcal{B}_{\bar{3}}\rangle+h.c..\label{lag2}
\end{eqnarray}
Here, $v=(1,\textbf{0})$ is the four velocity, $\rho_{ba}^{\mu}=ig_V{V}_{ba}^{\mu}/\sqrt{2}$, and $F^{\mu\nu}(\rho)=\partial^{\mu}\rho^{\nu}-\partial^{\nu}\rho^{\mu}
+\left[\rho^{\mu},\rho^{\nu}\right]$. $A_{\mu}$ and $\mathcal{V}_{\mu}$ stand for the axial current and vector current, respectively. They can be written as
\begin{eqnarray*}
A_{\mu} &=& \frac{1}{2}(\xi^{\dag}\partial_{\mu}\xi-\xi\partial_{\mu}\xi^{\dag})=\frac{i}{f_{\pi}}
\partial_{\mu}{P}+\ldots,\\
\mathcal{V}_{\mu} &=&
\frac{1}{2}(\xi^{\dag}\partial_{\mu}\xi+\xi\partial_{\mu}\xi^{\dag})
=\frac{i}{2f_{\pi}^2}\left[{P},\partial_{\mu}{P}\right]+\ldots,
\end{eqnarray*}
respectively. Here, $\xi=\text{exp}(i{P}/f_{\pi})$ and $f_{\pi}=132$ MeV. $\mathcal{B}_{\bar{3}}$ and $\mathcal{S}_{\mu} =-\sqrt{\frac{1}{3}}(\gamma_{\mu}+v_{\mu})\gamma^5\mathcal{B}_6+\mathcal{B}_{6\mu}^*$ denote the ground heavy baryons multiplets with their light quarks in the $\bar{3}$ and $6$ flavor representation, respectively. The matrices $\mathcal{B}_{\bar{3}}$, $\mathcal{B}_6$, ${P}$, and ${V}$ read as
\begin{eqnarray*}\left.\begin{array}{ll}
\mathcal{B}_{\bar{3}} = \left(\begin{array}{ccc}
        0    &\Lambda_c^+     \\
        -\Lambda_c^+       &0
\end{array}\right),
&\mathcal{B}_6 = \left(\begin{array}{ccc}
         \Sigma_c^{++}                  &\frac{\Sigma_c^{+}}{\sqrt{2}}\\
         \frac{\Sigma_c^{+}}{\sqrt{2}}      &\Sigma_c^{0}
\end{array}\right),\\
{P} = \left(\begin{array}{cc}
\frac{\pi^0}{\sqrt{2}}+\frac{\eta}{\sqrt{6}} &\pi^+  \nonumber\\
\pi^- &-\frac{\pi^0}{\sqrt{2}}+\frac{\eta}{\sqrt{6}}
\end{array}\right),
&{V} = \left(\begin{array}{cc}
\frac{\rho^0}{\sqrt{2}}+\frac{\omega}{\sqrt{2}} &\rho^+  \nonumber\\
\rho^- &-\frac{\rho^0}{\sqrt{2}}+\frac{\omega}{\sqrt{2}}
\end{array}\right).\end{array}\right.
\end{eqnarray*}

The effective Lagrangians describing the interactions between the strange mesons and light mesons are constructed in the $SU(3)$ symmetry \cite{Lin:1999ad,Nagahiro:2008mn}, i.e.,
\begin{eqnarray}
\mathcal{L}_{PPV} &=& \frac{ig}{2\sqrt{2}}\langle\partial^{\mu}P\left(PV_{\mu}-V_{\mu}P\right\rangle, \label{lag4}\\
\mathcal{L}_{VVP} &=& \frac{g_{VVP}}{\sqrt{2}}\epsilon^{\mu\nu\alpha\beta}
       \left\langle\partial_{\mu}V_{\nu}\partial_{\alpha}V_{\beta}P\right\rangle,\label{lag5}\\
\mathcal{L}_{VVV} &=& \frac{ig}{2\sqrt{2}}\langle\partial^{\mu}V^{\nu}\left(V_{\mu}V_{\nu}-V_{\nu}V_{\mu}\right)\rangle. \label{lag6}
\end{eqnarray}

After expanding Eqs. (\ref{lag1})-(\ref{lag6}), we can further obtain
\begin{eqnarray}
\mathcal{L}_{\sigma} &=& l_B\langle \bar{\mathcal{B}}_{\bar{3}}\sigma\mathcal{B}_{\bar{3}}\rangle
      -l_S\langle\bar{\mathcal{B}}_6\sigma\mathcal{B}_6\rangle,\\
\mathcal{L}_{{P}} &=&
        i\frac{g_1}{2f_{\pi}}\varepsilon^{\mu\nu\lambda\kappa}v_{\kappa}\langle\bar{\mathcal{B}}_6
        \gamma_{\mu}\gamma_{\lambda}\partial_{\nu}{P}\mathcal{B}_6\rangle\nonumber\\
      &&-\sqrt{\frac{1}{3}}\frac{g_4}{f_{\pi}}\langle\bar{\mathcal{B}}_6\gamma^5
      \left(\gamma^{\mu}+v^{\mu}\right)\partial_{\mu}{P}\mathcal{B}_{\bar{3}}\rangle+h.c.,\\
\mathcal{L}_{{V}} &=& \frac{1}{\sqrt{2}}\beta_Bg_V\langle\bar{\mathcal{B}}_{\bar{3}}v\cdot{V}\mathcal{B}_{\bar{3}}\rangle
   -\frac{\beta_Sg_V}{\sqrt{2}}\langle\bar{\mathcal{B}}_6v\cdot{V}\mathcal{B}_6\rangle\nonumber\\
    &&-\frac{\lambda_Ig_V}{\sqrt{6}}\varepsilon^{\mu\nu\lambda\kappa}v_{\mu}\langle \bar{\mathcal{B}}_6\gamma^5\gamma_{\nu}
        \left(\partial_{\lambda} {V}_{\kappa}-\partial_{\kappa} {V}_{\lambda}\right)\mathcal{B}_{\bar{3}}\rangle+h.c.\nonumber\\
        &&-i\frac{\lambda g_V}{3\sqrt{2}}\langle\bar{\mathcal{B}}_6\gamma_{\mu}\gamma_{\nu}
    \left(\partial^{\mu} {V}^{\nu}-\partial^{\nu} {V}^{\mu}\right)
    \mathcal{B}_6\rangle,\\
\mathcal{L}_{K^{(*)}K^{(*)}\sigma} &=& g_{\sigma }m_K\bar{K} K\sigma-g_{\sigma }m_{K^*}\bar{K}^{*}\cdot K^{*}\sigma,\\
\mathcal{L}_{P KK^*} &=& \frac{ig}{4}\left[\left(\bar{K}^{*\mu} K-\bar{K} K^{*\mu}\right)\left(\bm{\tau}\cdot\partial_{\mu}\bm{\pi}+\frac{\partial_{\mu}{\eta}}{\sqrt{3}}\right)\right.\nonumber\\
   &&\left.+\left(\partial_{\mu}\bar{K} K^{*\mu}-\bar{K}^{*\mu}\partial_{\mu}K\right)\left(\bm{\tau}\cdot\bm{\pi}+\frac{\eta}{\sqrt{3}}\right)\right],\\
\mathcal{L}_{{V} KK} &=& \frac{ig}{4}\left[\bar{K}\partial_{\mu}K
       -\partial_{\mu}\bar{K}K\right]\left(\bm{\tau}\cdot\bm{\rho}^{\mu}+{\omega}^{\mu}\right),\\
\mathcal{L}_{{V} K^*K^*} &=& \frac{ig}{4}
       \left[\left(\bar{K}_{\mu}^*\partial^{\mu}K^{*\nu}-\partial^{\mu}\bar{K}^{*\nu} K_{\mu}^*\right)\left(\bm{\tau}\cdot\bm{\rho}_{\nu}+\omega_{\nu}\right)\right.\nonumber\\
   &&\left.+\left(\partial^{\mu}\bar{K}^{*\nu}K_{\nu}^*-\bar{K}_{\nu}^*\partial^{\mu}K^{*\nu}\right)
       \left(\bm{\tau}\cdot\bm{\rho}_{\mu}+\omega_{\mu}\right)\right.\nonumber\\
       &&\left.+\left(\bar{K}_{\nu}^* K^*_{\mu}-\bar{K}_{\mu}^*K^*_{\nu}\right)
       \left(\bm{\tau}\cdot\partial^{\mu}\bm{\rho}^{\nu}+\partial^{\mu}\omega^{\nu}\right)\right],\\
\mathcal{L}_{P K^*K^*} &=& g_{VVP}\varepsilon_{\mu\nu\alpha\beta}
     \partial^{\mu}\bar{K}^{*\nu}\partial^{\alpha}K^{*\beta}\left(\bm{\tau}\cdot\bm{\pi}+\frac{\eta}{\sqrt{3}}\right),\\
\mathcal{L}_{V KK^*} &=& g_{VVP}\varepsilon_{\mu\nu\alpha\beta}
     \left(\partial^{\mu}\bar{K}^{*\nu}K+\bar{K}\partial^{\mu}{K}^{*\nu}\right)\nonumber\\
   &&\left(\bm{\tau}\cdot\partial^{\alpha}\bm{\rho}^{\beta}+\partial^{\alpha}{\omega}^{\beta}\right).
\end{eqnarray}

Coupling constants in the above Lagrangians are estimated with the quark model \cite{Liu:2011xc,Chen:2017xat}, $l_S=-2l_B=7.3$, $g_1=(\sqrt{8}/3)g_4=1.0$, $\beta_Sg_V=-2\beta_Bg_V=12.0$, $\lambda_Sg_V=-2\sqrt{2}\lambda_Ig_V=19.2~ \text{GeV}^{-1}$, $g_{\sigma}=-3.65$, and $g=12.00$. $g_{VVP}=3g^2/(32\sqrt{2}\pi^2f_{\pi})$ \cite{Kaymakcalan:1983qq}.

With these prepared effective Lagrangians, we can easily write down the scattering amplitudes for the $B_1M_2\to B_3M_4$ processes in the $t-$channel, where $B_1$ and $B_3$ stand for the initial and final baryons, respectively, and $M_2$ and $M_4$ stand for the inial and final mesons, respectively. The corresponding effective potentials can be related to the scattering amplitudes by the Breit approximation,
\begin{eqnarray}\label{breit}
\mathcal{V}_{E}^{B_1M_2\to B_3M_4}(\bm{q}) &=&
          -\frac{\mathcal{M}(B_1M_2\to B_3M_4)}
          {4\sqrt{m_{B_1}m_{M_2}m_{B_3}m_{M_4}}}.
\end{eqnarray}
Here, $m_{i}$ is the mass of the interaction hadron. $\mathcal{M}(B_1M_2\to B_3M_4)$ denotes the scattering amplitude for the $B_1M_2\to B_3M_4$ process by exchanging the light mesons ($\sigma$, $\pi$, $\eta$, $\rho$, and $\omega$). Next, we perform the Fourier transformation to obtain the effective potentials in the coordinate space $\mathcal{V}(\bm{r})$,
\begin{eqnarray}
\mathcal{V}_{E}(\bm{r}) =
          \int\frac{d^3\bm{q}}{(2\pi)^3}e^{i\bm{q}\cdot\bm{r}}
          \mathcal{V}_{E}(\bm{q})\mathcal{F}^2(q^2,m_E^2).\nonumber
\end{eqnarray}
In order to compensate the off-shell effect of the exchanged meson, we introduce a monopole form factor $\mathcal{F}(q^2,m_E^2)= (\Lambda^2-m_E^2)/(\Lambda^2-q^2)$ at every interactive vertex, where $\Lambda$, $m_E$, and $q$ are the cutoff parameter, the mass and four-momentum of the exchanged meson, respectively. In our numerical calculations, we vary the cutoff value in the range of $0.8\leq\Lambda\leq5.0$ GeV. According to the deuteron experience \cite{Tornqvist:1993ng,Tornqvist:1993vu}, the reasonable cutoff value is taken around 1.00 GeV. In the following discussion, the loosely bound state with the cutoff value around 1.00 GeV can be recommended as the prime hadronic molecular candidate.

For the $\Lambda_cK^{(*)}$ systems, the flavor wave function $|I,I_3\rangle$ can be expressed as $|1/2,1/2\rangle=|\Lambda_c^+K^{(*)+}\rangle$ and $|1/2,-1/2\rangle=|\Lambda_c^+K^{(*)0}\rangle$. For the $\Sigma_cK^{(*)}$ systems, their isospin $I$ can be taken as $1/2$ or $3/2$. The corresponding flavor wave functions $|I,I_3\rangle$ are
\begin{eqnarray*}&&\begin{array}{c}
\left|\frac{1}{2},\frac{1}{2}\right\rangle =
     \sqrt{\frac{2}{3}}\left|\Sigma_c^{++}{K}^{(*)0}\right\rangle
     -\frac{1}{\sqrt{3}}\left|\Sigma_c^{+}{K}^{(*)+}\right\rangle,\\
\left|\frac{1}{2},-\frac{1}{2}\right\rangle =
     \frac{1}{\sqrt{3}}\left|\Sigma_c^{+}{K}^{(*)0}\right\rangle
     -\sqrt{\frac{2}{3}}\left|\Sigma_c^{0}{K}^{(*)+}\right\rangle,
     \end{array}\\
&&\begin{array}{l}
\left|\frac{3}{2},\frac{3}{2}\right\rangle = \left|\Sigma_c^{++}{K}^{(*)+}\right\rangle,\\
\left|\frac{3}{2},\frac{1}{2}\right\rangle =
     \frac{1}{\sqrt{3}}\left|\Sigma_c^{++}{K}^{(*)0}\right\rangle+
     \sqrt{\frac{2}{3}}\left|\Sigma_c^{+}{K}^{(*)+}\right\rangle,\\
\left|\frac{3}{2},-\frac{1}{2}\right\rangle =\sqrt{\frac{2}{3}}
    \left|\Sigma_c^{+}{K}^{(*)0}\right\rangle
    +\frac{1}{\sqrt{3}}\left|\Sigma_c^{0}{K}^{(*)+}\right\rangle,\\
\left|\frac{3}{2},-\frac{3}{2}\right\rangle =
     \left|\Sigma_c^{0}{K}^{(*)0}\right\rangle,
     \end{array}
\end{eqnarray*}
respectively. When we consider the $S-D$ wave mixing effects, the spin-orbit wave functions $|{}^{2S+1}L_J\rangle$ are
\begin{eqnarray}\label{spinorbit}\left.\begin{array}{llll}
Y_cK[J^P=1/2^-]: &|{}^2S_{1/2}\rangle,\\
Y_cK^*[J^P=1/2^-]:   &|{}^{2}S_{1/2}\rangle,  &|{}^{4}D_{1/2}\rangle,\\
Y_cK^*[J^P=3/2^-]:   &|{}^{4}S_{3/2}\rangle,  &|{}^{2}D_{3/2}\rangle,  &|{}^{4}D_{3/2}\rangle.
\end{array}\right.\end{eqnarray}
The general expressions of the spin-orbit wave functions $|{}^{2S+1}L_J\rangle$ for the $Y_cK^{(*)}$ systems read as
\begin{eqnarray}
Y_cK:\, \left|{}^{2S+1}L_{J}\right\rangle &=& \sum_{m_S,m_L}C^{J,M}_{\frac{1}{2}m_S,Lm_L}\chi_{\frac{1}{2}m}|Y_{L,m_L}\rangle,\nonumber\\
Y_c{K}^*: \left|{}^{2S+1}L_{J}\right\rangle &=&
\sum_{m,m'}^{m_S,m_L}C^{S,m_S}_{\frac{1}{2}m,1m'}C^{J,M}_{Sm_S,Lm_L}
          \chi_{\frac{1}{2}m}\epsilon^{m'}|Y_{L,m_L}\rangle.\nonumber
\end{eqnarray}
Here, $C^{J,M}_{\frac{1}{2}m_S,Lm_L}$, $C^{S,m_S}_{\frac{1}{2}m,1m'}$, and $C^{J,M}_{Sm_S,Lm_L}$ are the Clebsch-Gordan coefficients. $\chi_{\frac{1}{2}m}$ and $Y_{L,m_L}$ stand for the spin wave function and the spherical harmonics function, respectively. $\epsilon$ is the polarization vector for the vector meson with $\epsilon_{\pm}^{m}=\mp\frac{1}{\sqrt{2}}\left(\epsilon_x^{m}{\pm}i\epsilon_y^{m}\right)$ and $\epsilon_0^{m}=\epsilon_z^{m}$, which satisfies $\epsilon_{\pm1}= \frac{1}{\sqrt{2}}\left(0,\pm1,i,0\right)$ and $\epsilon_{0} =\left(0,0,0,-1\right)$.

\section{The OBE effective potentials and the numerical results}\label{sec3}

Following the above procedures, we can deduce the concrete OBE effective potentials for the $Y_cK^{(*)}$ systems with different quantum configurations. After that, we adopt the obtained OBE effective potentials to solve the coupled channel Schr\"{o}dinger equations. By doing this, we can search for the bound state solutions. A system with the reasonable bound state solutions can be recommended as the good hadronic molecular candidate, where the binding energy is taken from several MeV to several tens MeV, and the root-mean-square (RMS) radius is a few fm or larger.


\subsection{The $\Lambda_cK^{(*)}$ systems}

The total OBE effective potentials for the single $\Lambda_cK$ system can be written as
\begin{eqnarray}
V_{\Lambda_cK\to\Lambda_cK} &=& l_Bg_{\sigma}\chi_3^{\dag}\chi_1Y(\Lambda,m_{\sigma},r)\nonumber\\
    &&+\frac{\beta_Bg_Vg}{4}\chi_3^{\dag}\chi_1Y(\Lambda,m_{\omega},r).
    \label{lamk}
\end{eqnarray}
Here, we define
\begin{eqnarray}
Y(\Lambda,m,{r}) &=& \frac{1}{4\pi r}(e^{-mr}-e^{-\Lambda
r})-\frac{\Lambda^2-m^2}{8\pi \Lambda}e^{-\Lambda r}.
\end{eqnarray}
As shown in Eq. (\ref{lamk}), there exist the $\sigma$ exchange and $\omega$ exchange interactions, which contribute in the intermediate range and the short range, respectively. The $\sigma$ exchange provides an attractive interaction, whereas the $\omega$ exchange interaction is repulsive. Here, the $\rho$ exchange interaction is strongly suppressed as the isospin forbidden in the $\Lambda_c-\Lambda_c-\rho$ coupling. Since the $KK\pi(\eta)$ coupling is forbidden by the spin-parity conservation, the pseudoscalr meson $(\pi/\eta)$ exchanges interactions are strongly suppressed, either.

After solving the Schr\"{o}dinger equation, we don't find bound state solutions in the cutoff region $0.8\leq\Lambda\leq5.0$ GeV. Thus, the OBE effective potentials for the $\Lambda_cK$ system is not strong enough to bind a bound state.

For the single $S-$wave $\Lambda_cK^*$ systems with $J^P=1/2^-$ and $3/2^-$, their OBE effective potentials are the same, i.e.,
\begin{eqnarray}
V_{\Lambda_cK^*\to\Lambda_cK^*} &=& l_Bg_{\sigma}(\bm{\epsilon}_2\cdot\bm{\epsilon}_4^{\dag})\chi_3^{\dag}\chi_1Y(\Lambda,m_{\sigma},r)\nonumber\\
    &&+\frac{\beta_Bg_Vg}{4}(\bm{\epsilon}_2\cdot\bm{\epsilon}_4^{\dag})\chi_3^{\dag}\chi_1Y(\Lambda,m_{\omega},r).
\end{eqnarray}
When we consider the $S-D$ wave mixing effects, the operator $\bm{\epsilon}_2\cdot\bm{\epsilon}_4^{\dag}$ will be replaced by the unit matrix $\mathcal{I}=\langle{}^{2S'+1}L'_{J'}|\bm{\epsilon}_2\cdot\bm{\epsilon}_4^{\dag}|{}^{2S+1}L_J\rangle$ in the numerical calculations, which indicates the OBE effective potentials are the exactly the same with those for the $\Lambda_cK$ system with $1/2^-$. In the cutoff region $0.8\leq\Lambda\leq5.0$ GeV, we cannot find the bound state solutions, either.

In this work, we further perform the coupled channel analysis on the $\Lambda_cK^*/\Sigma_cK^*$ interactions, the corresponding OBE effective potentials are
\begin{eqnarray}
V_{\Lambda_cK^*}^{\text{C}} &=& \left(\begin{array}{cc}
V_{\Lambda_cK^*\to\Lambda_cK^*}    &V_{\Sigma_cK^*\to\Lambda_cK^*}\\
V_{\Lambda_cK^*\to\Sigma_cK^*}     &V_{\Sigma_cK^*\to\Sigma_cK^*}\end{array}\right),
\end{eqnarray}
with
\begin{eqnarray}
V_{\Lambda_cK^*\to\Sigma_cK^*} &=& \frac{1}{6}\frac{g_4g_{VVP}}{f_{\pi}}
\mathcal{F}_1(r,\bm{\sigma},i\bm{\epsilon}_2\times\bm{\epsilon}_4^{\dag})
     Y(\Lambda_0,m_{\pi0},r)\nonumber\\
     &&-\frac{1}{6\sqrt{2}}\frac{\lambda_Ig_Vg}{m_{K^*}}
     \mathcal{F}_2(r,\bm{\sigma},i\bm{\epsilon}_2\times\bm{\epsilon}_4^{\dag})
     Y(\Lambda_0,m_{\rho0},r),\label{lamks1}\nonumber\\\\
V_{\Sigma_cK^*\to\Sigma_cK^*} &=& \frac{1}{2}l_Sg_{\sigma}\chi_3^{\dag}\chi_1
     \bm{\epsilon}_2\cdot\bm{\epsilon}_3^{\dag}Y(\Lambda,m_{\sigma},r)\nonumber\\
     &&-\frac{g_1g_{VVP}}{6\sqrt{2}f_{\pi}}
     \mathcal{F}_1(r,\bm{\sigma},i\bm{\epsilon}_2\times\bm{\epsilon}_4^{\dag})
     \mathcal{G}(I)Y(\Lambda,m_{\pi},r)\nonumber\\
     &&-\frac{g_1g_{VVP}}{18\sqrt{2}f_{\pi}}
     \mathcal{F}_1(r,\bm{\sigma},i\bm{\epsilon}_2\times\bm{\epsilon}_4^{\dag})
     Y(\Lambda,m_{\eta},r)\nonumber\\
     &&+\frac{1}{8}\beta_Sg_Vg\chi_3^{\dag}\chi_1
     \bm{\epsilon}_2\cdot\bm{\epsilon}_3^{\dag}\mathcal{G}(I)Y(\Lambda,m_{\rho},r)\nonumber\\
     &&+\frac{\lambda_Sg_Vg}{8\sqrt{3}m_{\Sigma_c}}\chi_3^{\dag}\chi_1
     \bm{\epsilon}_2\cdot\bm{\epsilon}_3^{\dag}\mathcal{G}(I)\nabla^2Y(\Lambda,m_{\rho},r)\nonumber\\
     &&-\frac{\lambda_Sg_Vg}{24\sqrt{3}m_{K^*}}
     \mathcal{F}_2(r,\bm{\sigma},i\bm{\epsilon}_2\times\bm{\epsilon}_4^{\dag})
     \mathcal{G}(I)Y(\Lambda,m_{\rho},r)\nonumber\\
     &&+\frac{1}{8}\beta_Sg_Vg\chi_3^{\dag}\chi_1
     \bm{\epsilon}_2\cdot\bm{\epsilon}_3^{\dag}Y(\Lambda,m_{\omega},r)\nonumber\\
     &&+\frac{\lambda_Sg_Vg}{8\sqrt{3}m_{\Sigma_c}}\chi_3^{\dag}\chi_1
     \bm{\epsilon}_2\cdot\bm{\epsilon}_3^{\dag}\nabla^2Y(\Lambda,m_{\omega},r)\nonumber\\
     &&-\frac{\lambda_Sg_Vg}{24\sqrt{3}m_{K^*}}
     \mathcal{F}_2(r,\bm{\sigma},i\bm{\epsilon}_2\times\bm{\epsilon}_4^{\dag})
     Y(\Lambda,m_{\omega},r).\label{SigKx}
\end{eqnarray}
Here, the value of the isospin factor $\mathcal{G}(I)$ is taken as $\mathcal{G}(I=1/2)=-2$, $\mathcal{G}(I=3/2)=1$. The variables in Eq. (\ref{lamks1}) are $\Lambda_0^2 =\Lambda^2-q_0^2$, $m_{{\pi0}}^2=m_{\pi}^2-q_0^2$, $m_{\rho0}^2=m_{\rho}^2-q_0^2$ with $q_0 =
\frac{M_{\Sigma_c}^2-M_{\Lambda_c}^2}{2(M_{\Sigma_c}+M_{K^*})}$. And we define useful operators, i.e.,
\begin{eqnarray}
\mathcal{F}_1(r,\bm{a},\bm{b}) &=& \chi_3^{\dag}
     \left(\bm{a}\cdot\bm{b}\nabla^2
     +S(\hat{r},\bm{a},\bm{b})
     r\frac{\partial}{\partial r}\frac{1}{r}\frac{\partial}{\partial r}\right)\chi_1,\\
\mathcal{F}_2(r,\bm{a},\bm{b}) &=& \chi_3^{\dag}
     \left(2\bm{a}\cdot\bm{b}\nabla^2
     -S(\hat{r},\bm{a},\bm{b})
     r\frac{\partial}{\partial r}\frac{1}{r}\frac{\partial}{\partial r}\right)\chi_1.
\end{eqnarray}
Here, the $\bm{a}\cdot\bm{b}$ and $S(\hat{r},\bm{a},\bm{b})$ stand for the spin-spin interactions and the tensor force operators, respectively. The corresponding matrices elements can be obtained by sandwiched between the spin-orbit wave functions as presented in the Eq. (\ref{spinorbit}), i.e.,
\begin{eqnarray}
i\bm{\sigma}\cdot(\bm{\epsilon}_2\times\bm{\epsilon}_4^{\dag})&\mapsto&
\left\{\begin{array}{ll}\left(\begin{array}{cc}-2  &0\\ 0 &1\end{array}\right),  &J^P=1/2^-\\
\left(\begin{array}{ccc}1  &0  &0\\0  &-2  &0\\ 0  &0  &1\end{array}\right),  &J^P=3/2^-\end{array}\right.\\
S(\hat{r},\bm{\sigma},i\bm{\epsilon}_2\times\bm{\epsilon}_4^{\dag})&\mapsto&
\left\{\begin{array}{ll}\left(\begin{array}{cc}0  &-\sqrt{2}\\ -\sqrt{2} &-2\end{array}\right),  &J^P=1/2^-\\
\left(\begin{array}{ccc}0  &1  &2\\ 1  &0  &-1\\ 2  &-1  &0\end{array}\right),  &J^P=3/2^-\end{array}\right.
\end{eqnarray}

\renewcommand\tabcolsep{0.2cm}
\renewcommand{\arraystretch}{1.7}
\begin{table*}[!hbtp]
\caption{The bound state solutions (the binding energy $E$, the root-mean-square radius $r_{RMS}$, and the probabilities $P_i(\%)$ for all the discussed channels) for the coupled $\Lambda_cK^*/\Sigma_cK^*$ systems with $I(J^P)=1/2(1/2^-)$ and $1/2(3/2^-)$. Here, $E$, $r_{RMS}$, and $\Lambda$ are in units of MeV, fm, and GeV, respectively. The dominant channels are labeled in a bold manner.}\label{num1}
\begin{tabular}{c|ccc|cccccc}
\toprule[1pt]
 $I(J^{P})$      &$\Lambda$    &$E$       &$r_{RMS}$
      &$\Lambda_cK^*({}^2S_{1/2})$   &$\Lambda_cK^*({}^{4}D_{1/2})$     &$\Sigma_cK^*({}^{2}S_{1/2})$      &$\Sigma_cK^*({}^{4}D_{1/2})$\\\midrule[1pt]
$1/2(1/2^-)$    &1.56   &$-0.14$   &6.11  &\bf{98.82}  &$\sim0$   &1.14   &0.04\\
                &1.58   &$-2.14$   &2.62  &\bf{97.11}  &0.01    &2.82   &0.06\\
                &1.60   &$-6.02$   &1.56  &\bf{95.12}   &0.02    &4.79   &0.07\\
                &1.62   &$-11.57$  &1.12  &\bf{93.18}   &0.03    &6.72  &0.07\\
                \midrule[1pt]
$I(J^{P})$      &$\Lambda$    &$E$       &$r_{RMS}$
      &$\Lambda_cK^*({}^4S_{3/2})$   &$\Lambda_cK^*({}^{2}D_{3/2})$     &$\Lambda_cK^*({}^{4}D_{3/2})$   &$\Sigma_cK^*({}^{4}S_{3/2})$      &$\Sigma_cK^*({}^{2}D_{3/2})$   &$\Sigma_cK^*({}^{4}D_{3/2})$
      \\\midrule[1pt]
$1/2(3/2^-)$  &1.34  &$-0.07$  &6.35 &\bf{94.23}  &0.03  &0.11 &4.89  &0.22 &0.52\\
              &1.36  &$-3.06$  &2.07 &\bf{84.54}  &0.08  &0.28 &13.36  &0.53 &1.20\\
              &1.38   &$-8.93$  &1.18 &\bf{75.92}   &0.12  &0.42  &21.07  &0.76 &1.71\\
              &1.40   &$-16.80$ &0.87  &\bf{69.12}   &0.14  &0.52  &27.27  &0.91 &2.05\\
\bottomrule[1pt]
\end{tabular}
\end{table*}

With these deduced effective potentials, we search for the bound state solutions for the coupled $\Lambda_cK^*/\Sigma_cK^*$ systems in the cutoff range $0.8\leq\Lambda\leq5.0$ GeV. In Table \ref{num1}, we collect the corresponding numerical results, which include the cutoff dependence of the binding energy $E$, the root-mean-square radius $r_{RMS}$, and the probabilities $P_i(\%)$ for all the discussed channels.

For the coupled $\Lambda_cK^*/\Sigma_cK^*$ system with $I(J^P)=1/2(1/2^-)$, there exist four channels, the $\Lambda_cK^*({}^2S_{1/2}, {}^4D_{1/2})$ channels and the $\Sigma_cK^*({}^2S_{1/2}, {}^4D_{1/2})$ channels after considering both the $S-D$ wave mixing effects and the coupled channel effects. As presented in Table \ref{num1}, when the cutoff is taken as 1.56 GeV, the binding energy is $-0.14$ MeV, the RMS radius is 6.11 fm, and the probability for the $\Lambda_cK^*({}^2S_{1/2})$ channel is 98.82\%. As the cutoff $\Lambda$ increases to 1.62 GeV, the binding energy becomes $-11.57$ MeV, the RMS radius becomes 1.12 fm, and the $\Lambda_cK^*({}^2S_{1/2})$ is still the dominant channel with the probability around 93.18\%. From the current numerical results, in the cutoff range around 1.60 GeV, we can obtain the weakly bound state with the reasonable loosely bound state solutions, and the dominant channel is the $\Lambda_cK^*({}^2S_{1/2})$ with its probability over 90\%. Since the cutoff value is close to the empirical value $\Lambda\sim 1.00$ GeV for the deuteron \cite{Tornqvist:1993ng,Tornqvist:1993vu}, we conclude that the coupled $\Lambda_cK^*/\Sigma_cK^*$ systems with $I(J^P)=1/2(1/2^-)$ can be recommended as a good hadronic molecular candidate.

For the coupled $\Lambda_cK^*/\Sigma_cK^*$ system with $I(J^P)=1/2(3/2^-)$, there include the $\Lambda_cK^*({}^4S_{3/2}, {}^2D_{3/2}, {}^4D_{3/2})$ channels and the $\Sigma_cK^*({}^4S_{3/2}, {}^2D_{3/2}, {}^4D_{3/2})$ channels when we consider both the coupled channel effects and the $S-D$ wave mixing effects. As shown in Table \ref{num1}, we can obtain loosely bound state solutions at the cutoff larger than 1.34 GeV, where the binding energy is from several to ten MeV, and the RMS radius is larger than 1.00 fm, the dominant channel is the $\Lambda_cK^*({}^4S_{3/2})$ channel. As the increasing of the cutoff value, the $\Sigma_cK^*({}^4S_{3/2})$ channel becomes more and more important, when the cutoff is 1.40 GeV, the probability of the $S-$wave $\Sigma_cK^*$ component turns into 27.27\%. If we still adopt the experience of the deuteron~\cite{Tornqvist:1993ng,Tornqvist:1993vu}, the coupled $\Lambda_cK^*/\Sigma_cK^*$ system with $1/2(3/2^-)$ can be a good hadronic molecular candidate, it is mainly composed by the $\Lambda_cK^*({}^4S_{3/2})$ channel, followed by the $\Sigma_cK^*({}^4S_{3/2})$ channel.

In addition, we find that the coupled channel effects play an important role in forming these $\Lambda_cK^*/\Sigma_cK^*$ bound states with $1/2(1/2^-, 3/2^-)$, since there don't exist bound state solutions in the single $\Lambda_cK^*$ systems.

\subsection{The $\Sigma_cK^{(*)}$ systems}

The OBE effective potentials for the single $\Sigma_cK$ is
\begin{eqnarray}
V_{\Sigma_cK\to\Sigma_cK} &=& \frac{1}{2}l_Sg_{\sigma}\chi_3^{\dag}\chi_1Y(\Lambda,m_{\sigma},r)\nonumber\\
    &&+\frac{\mathcal{G}(I)}{8}\beta_Sg_Vg\chi_3^{\dag}\chi_1Y(\Lambda,m_{\rho},r)\nonumber\\
    &&-\frac{\mathcal{G}(I)}{24m_{\Sigma_c}}\lambda_Sg_Vg\chi_3^{\dag}\chi_1
    \nabla^2Y(\Lambda,m_{\rho},r)\nonumber\\
    &&+\frac{1}{8}\beta_Sg_Vg\chi_3^{\dag}\chi_1Y(\Lambda,m_{\omega},r)\nonumber\\
    &&-\frac{1}{24m_{\Sigma_c}}\lambda_Sg_Vg\chi_3^{\dag}\chi_1\nabla^2Y(\Lambda,m_{\omega},r).
\end{eqnarray}
Here, there exists the extra $\rho$ exchange interaction in comparison to the $\Lambda_cK$ system, and it provides the attractive and repulsive forces for the $\Sigma_cK$ system with $I=1/2$ and $3/2$, respectively. Therefore, it is possible to find the bound state solutions for the $\Sigma_cK$ system with $I=1/2$ as the stronger attractive OBE effective potentials. After solving the coupled channel Schr\"{o}dinger equation, our results show that there exist no bound state solutions for the iso-quartet $\Sigma_cK$ system. For the iso-doublet $\Sigma_cK$ system, as presented in Table \ref{num2}, we can obtain the reasonable loosely bound state solutions when the cutoff $\Lambda$ is larger than 2.00 GeV.

\renewcommand\tabcolsep{0.2cm}
\renewcommand{\arraystretch}{1.6}
\begin{table*}[!hbtp]
\caption{The bound state solutions (the binding energy $E$, the root-mean-square radius $r_{RMS}$, and the probabilities $P_i(\%)$ for all the discussed channels) for the single $\Sigma_cK$ and the coupled $\Sigma_cK/\Lambda_cK^*/\Sigma_cK^*$ systems with $I(J^P)=1/2(1/2^-)$ and $3/2(1/2^-)$. Here, $E$, $r_{RMS}$, and $\Lambda$ are in units of MeV, fm, and GeV, respectively. The dominant channels are labeled in a bold manner.} \label{num2}
\begin{tabular}{c|ccc|cccccccc}
\toprule[1pt]\toprule[1pt]
&\multicolumn{3}{c|}{Single channel}    &\multicolumn{8}{c}{Coupled channel}\\\midrule[1pt]
  $I(J^P)$ &$\Lambda$   &$E$ &$r_{RMS}$
 &$\Lambda$    &$E$  &$r_{RMS}$
      &$\Sigma_cK({}^2S_{1/2})$    &$\Lambda_cK^*({}^2S_{1/2})$   &$\Lambda_cK^*({}^{4}D_{1/2})$     &$\Sigma_cK^*({}^{2}S_{1/2})$      &$\Sigma_cK^*({}^{4}D_{1/2})$\\\hline
 $1/2(1/2^-)$
 &2.00    &$-0.94$    &4.78
     &0.90   &$-0.36$    &6.14    &\bf{98.85}   &0.61    &0.47      &0.01     &0.06\\
 &2.20    &$-4.80$    &2.44
     &0.95   &$-3.28$    &3.04    &\bf{97.61}   &1.34    &0.92      &0.02     &0.12\\
 &2.40    &$-10.96$    &1.68
     &1.00   &$-9.27$    &1.91    &\bf{96.11}   &2.25    &1.42      &0.04     &0.18\\
 &2.60    &$-18.92$    &1.31
     &1.05   &$-18.44$   &1.42    &\bf{94.60}   &3.18    &1.92      &0.06     &0.24\\\hline
 $3/2(1/2^-)$
 &\ldots    &\ldots    &\ldots      &1.28   &$-2.58$    &2.85
        &\bf{92.77}      &\ldots      &\ldots     &7.11    &0.12\\
 &\ldots    &\ldots    &\ldots      &1.29   &$-12.82$    &1.23
        &\bf{85.78}      &\ldots      &\ldots     &13.99    &0.22\\
 &\ldots    &\ldots    &\ldots      &1.30   &$-28.46$    &0.80
        &\bf{80.03}      &\ldots      &\ldots     &19.67    &0.30\\
 &\ldots    &\ldots    &\ldots      &1.31   &$-48.10$   &0.61
        &\bf{75.36}      &\ldots      &\ldots     &24.29    &0.35\\
 \bottomrule[1pt]
\bottomrule[1pt]
\end{tabular}
\end{table*}

When we further perform the coupled $\Sigma_cK/\Lambda_cK^*/\Sigma_cK^*$ analysis. There can allow the $\pi-$exchange interactions for both of the $\Lambda_cK^*\to\Sigma_cK$ and $\Sigma_cK^*\to\Sigma_cK$ process, which plays very important role in binding the deuteron. The corresponding OBE effective potentials can be expressed as
\begin{eqnarray}
V_{\Sigma_cK}^{\text{C}} &=& \left(\begin{array}{ccc}
V_{\Sigma_cK\to\Sigma_cK}    &V_{\Lambda_cK^*\to\Sigma_cK}    &V_{\Sigma_cK^*\to\Sigma_cK}\\
V_{\Sigma_cK\to\Lambda_cK^*}     &V_{\Lambda_cK^*\to\Lambda_cK^*} &V_{\Sigma_cK^*\to\Lambda_cK^*}\\
V_{\Sigma_cK\to\Sigma_cK^*}     &V_{\Lambda_cK^*\to\Sigma_cK^*} &V_{\Sigma_cK^*\to\Sigma_cK^*}\end{array}\right),
\end{eqnarray}
with
\begin{eqnarray}
V_{\Lambda_cK^*\to\Sigma_cK} &=& -\frac{1}{6}\frac{g_4g}{f_{\pi}\sqrt{m_Km_{K^*}}}
    \mathcal{F}_1(r,\bm{\sigma},\bm{\epsilon}_2)
     U(\Lambda_1,m_{\pi1},r)\nonumber\\
     && -\frac{\lambda_Ig_Vg_{VVP}}{3\sqrt{2}}\sqrt{\frac{m_{K^*}}{m_K}}
    \mathcal{F}_2(r,\bm{\sigma},\bm{\epsilon}_2)
     Y(\Lambda_1,m_{\rho1},r),\label{LamKx1}\nonumber\\\\
V_{\Sigma_cK^*\to\Sigma_cK} &=& \frac{g_1g\mathcal{F}_1(r,\bm{\sigma},\bm{\epsilon}_2)}
  {24\sqrt{2}f_{\pi}\sqrt{m_Km_{K^*}}}
     \mathcal{G}(I)Y(\Lambda_2,m_{\pi2},r)\nonumber\\
     &&+\frac{g_1g}{72\sqrt{2}f_{\pi}\sqrt{m_Km_{K^*}}}
    \mathcal{F}_1(r,\bm{\sigma},\bm{\epsilon}_2)
     Y(\Lambda_2,m_{\eta2},r)\nonumber\\
     &&+\frac{\lambda_Sg_Vg_{VVP}}{6\sqrt{3}}\sqrt{\frac{m_{K^*}}{m_K}}
    \mathcal{F}_2(r,\bm{\sigma},\bm{\epsilon}_2)
     \mathcal{G}(I)Y(\Lambda_2,m_{\rho2},r)\nonumber\\
     &&+\frac{\lambda_Sg_Vg_{VVP}}{6\sqrt{3}}\sqrt{\frac{m_{K^*}}{m_K}}
    \mathcal{F}_2(r,\bm{\sigma},\bm{\epsilon}_2)
     Y(\Lambda_2,m_{\omega2},r).\label{SigK1}\nonumber\\
\end{eqnarray}
Here, we define an useful function in Eq. (\ref{LamKx1}), i.e.,
\begin{eqnarray*}
U(\Lambda,m,{r}) &=& \frac{1}{4\pi r}\left(\cos(mr)-e^{-\Lambda r}\right)-\frac{\Lambda^2+m^2}{8\pi \Lambda}e^{-\Lambda r}.
\end{eqnarray*}
The variables in the above effective potentials (\ref{LamKx1})$-$(\ref{SigK1}) are defined as
$q_1=\frac{M_{\Lambda_c}^2+M_{K}^2-M_{\Sigma_c}^2-M_{K^*}^2}{2(M_{\Sigma_c}+M_{K})}$, $\Lambda_1^2=\Lambda^2-q_1^2$, $m_{\pi1}^2=q_1^2-m_{\pi}^2$, $m_{\rho1}^2=m_{\rho}^2-q_1^2$, $q_2=\frac{M_{K^*}^2-M_{K}^2}{2(M_{\Sigma_c}+M_{K})}$, $\Lambda_2^2=\Lambda^2-q_2^2$, $m_{\pi2}^2=m_{\pi}^2-q_2^2$, $m_{\eta2}^2=m_{\eta}^2-q_2^2$, $m_{\rho2}^2=m_{\rho}^2-q_2^2$, $m_{\omega2}^2=m_{\omega}^2-q_2^2$. After considering the $S-D$ wave mixing effects, the matrix elements for the spin-spin interaction and tensor force operators read as $\bm{\sigma}\cdot\bm{\epsilon_2}\mapsto (\begin{array}{cc}\sqrt{3} &0\end{array} )$ and $ S(\hat{r},\bm{\sigma},\bm{\epsilon_2})\mapsto (\begin{array}{cc}0 &-\sqrt{6}\end{array} )$, respectively.

In Table \ref{num2}, we collect the bound state solutions (the binding energy $E$, the root-mean-square radius $r_{RMS}$, and the probabilities $P_i(\%)$ for all the discussed channels) for the coupled $\Sigma_cK/\Lambda_cK^*/\Sigma_cK^*$ systems with $I(J^P)=0,1(1/2^-)$.

For the $\Sigma_cK/\Lambda_cK^*/\Sigma_cK^*$ system with $I(J^P)=1/2(1/2^-)$, there exist the $\Sigma_cK({}^2S_{1/2})$ channel, the $\Lambda_cK^*({}^2S_{1/2},{}^4D_{1/2})$ channels, and the $\Sigma_cK^*({}^2S_{1/2},{}^4D_{1/2})$ channels. The reasonable loosely bound state solutions emerge at the cutoff $\Lambda=0.90$ GeV, where the binding energy is $-0.36$ MeV, the RMS radius is 4.78 fm, and the dominant channel is the $\Sigma_cK({}^2S_{1/2})$ with the probability $P=98.85$\%. When the cutoff increases to 1.05 GeV, this bound state binds deeper, the binding energy is $-18.44$ MeV, the RMS radius decreases to 1.42 fm, and the $\Sigma_cK({}^2S_{1/2})$ channel is still the dominant channel with its probability around 95\%. For the remaining channels, their probabilities are very tiny. Compared to the bound state properties in the single channel case, the cutoff is very close to the reasonable value $\Lambda\sim$1.00 GeV. Therefore, the coupled $\Sigma_cK/\Lambda_cK^*/\Sigma_cK^*$ system with $I(J^P)=1/2(1/2^-)$ can be the prime molecular candidate, and the coupled channel effects play an important role for the formation of this bound state.

For the $\Sigma_cK/\Sigma_cK^*$ system with $I(J^P)=3/2(1/2^-)$, there include the $\Sigma_cK({}^2S_{1/2})$ channel and the $\Sigma_cK^*({}^2S_{1/2},{}^4D_{1/2})$ channels. We find a weakly bound state at the cutoff $\Lambda=$1.28 GeV, the binding energy is $E=-2.58$ MeV, the RMS radius is $r_{RMS}=$2.85 fm, and the dominant channel is the $\Sigma_cK({}^2S_{1/2})$ with the probability $P=92.77$\%. With the increasing of the cutoff value, the $\Sigma_cK^*$ channel becomes more and more important. As the cutoff increases to 1.31 GeV, the binding energy turns into $-48.10$ MeV, the RMS radius decreases to 0.61 fm, and the probability for the $\Sigma_cK^*({}^2S_{1/2})$ is 24.29\%. However, the binding energy depends very sensitively with the cutoff. Thus, we cannot draw a definite conclusion that the $\Sigma_cK/\Sigma_cK^*$ system with $I(J^P)=3/2(1/2^-)$ as a good hadronic molecular candidate.

\renewcommand\tabcolsep{0.1cm}
\renewcommand{\arraystretch}{1.6}
\begin{table}[!hbtp]
\caption{The $\Lambda$ dependence of the obtained bound-state solutions (the binding energy $E$ and the root-mean-square radius $r_{RMS}$) for the single $\Sigma_c K^*$ systems. Here, $E$, $r_{RMS}$, and $\Lambda$ are in units of MeV, fm, and GeV, respectively.}\label{num3}
\begin{tabular}{cccc|cccc}
\toprule[1pt]\toprule[1pt]
$I(J^P)$ &$\Lambda$   &$E$ &$r_{RMS}$
 &$I(J^P)$  &$\Lambda$    &$E$  &$r_{RMS}$ \\\hline
 $1/2(1/2^-)$   &1.70     &$-0.50$   &5.32
 &$1/2(3/2^-)$   &0.88     &$ -0.25$   &6.06\\
                &2.00     &$-3.32$   &2.64
                &&0.98     &$-3.32$   &2.58\\
                &2.30     &$-8.31$   &1.81
                &&1.08     &$-10.72$   &1.59\\
                &2.60     &$-15.30$   &1.42
                &&1.18     &$-23.27$   &1.15\\\hline
 $3/2(1/2^-)$   &1.28     &$-0.11$   &6.24
 &$3/2(3/2^-)$  &\ldots    &\ldots    &\ldots\\
                &1.31     &$-2.42$   &2.50
                &&\ldots    &\ldots    &\ldots\\
                &1.34     &$-7.78$   &1.43
                &&\ldots    &\ldots    &\ldots\\
                &1.37     &$-16.71$   &1.00
                &&\ldots    &\ldots    &\ldots\\
 \bottomrule[1pt]
\bottomrule[1pt]
\end{tabular}
\end{table}

For the $\Sigma_cK^*$ systems, the isospin and spin-parity configurations $I(J^P)$ include $1/2(1/2^-)$, $1/2(3/2^-)$, $3/2(1/2^-)$, and $3/2(3/2^-)$ after considering the $S-D$ wave mixing effects. The relevant OBE effective potentials are presented in Eq. (\ref{SigKx}). Our results indicate that there exist the reasonable loosely bound state solutions for the $\Sigma_cK^*$ states with $I(J^P)=1/2(1/2^-)$, $1/2(3/2^-)$, and $3/2(1/2^-)$ in the cutoff range $0.80\leq\Lambda\leq5.00$ GeV. As shown in Table \ref{num3}, for the $\Sigma_cK^*$ systems with $1/2(3/2^-)$ and $3/2(1/2^-)$, the binding energy around several to several tens MeV and the RMS radius around several fm appear at the cutoff around 1.00 GeV, which is comparable to the value in the deuteron \cite{Tornqvist:1993ng,Tornqvist:1993vu}. Therefore, these two states can be suggested as the good hadronic molecular candidates. For the $\Sigma_cK^*$ system with $1/2(1/2^-)$, the loosely bound state solutions appear as the cutoff is larger than 1.70 GeV, which is slightly far away from the empirical value for the deuteron \cite{Tornqvist:1993ng,Tornqvist:1993vu}, in this work, we cannot exclude that the $\Sigma_cK^*$ system with $1/2(1/2^-)$ as a suitable molecular candidate.

In summary, our results can predict several possible open charm molecular pentaquarks, the coupled $\Lambda_cK^*/\Sigma_cK^*$ molecular states with $I(J^P)=1/2(1/2^-,3/2^-)$, the coupled $\Sigma_cK/\Lambda_cK^*/\Sigma_cK^*$ molecular states with $I(J^P)=1/2(1/2^-)$, and the single $\Sigma_cK^*$ states with $I(J^P)=1/2(1/2^-,3/2^-)$ and $3/2(1/2^-)$. And the coupled channel effects do play the very important role in generating these coupled channel molecular candidates.

The study of the strong decay behaviors is very helpful to the search of these predicted open flavor molecular pentaquarks. According to the conservation of the quantum numbers and the limit of the phase space, we collect the important strong decay decay channels as follows, i.e.,
\begin{eqnarray}
\Sigma_cK/\Lambda_cK^*/\Sigma_cK^*[1/2(1/2^-)] &\to& \left\{D_sN, \Lambda_cK\right\},\nonumber\\
\Lambda_cK^*/\Sigma_cK^*[1/2(1/2^-)] &\to& \left\{D_s^{(*)}N, \Lambda_cK, \Sigma_cK\right\},\nonumber\\
\Lambda_cK^*/\Sigma_cK^*[1/2(3/2^-)] &\to& \left\{D_s^{*}N\right\},\nonumber\\
\Sigma_cK^*[1/2(1/2^-)] &\to& \left\{D_s^{(*)}N, \Lambda_cK^{(*)}, \Sigma_cK\right\},\nonumber\\
\Sigma_cK^*[1/2(3/2^-)] &\to& \left\{D_s^{*}N, \Lambda_cK^{*}, \Sigma_c^*K\right\},\nonumber\\
\Sigma_cK^*[3/2(1/2^-)] &\to& \left\{D_s^{*}\Delta, \Sigma_cK\right\}.\nonumber
\end{eqnarray}

\subsection{The predictions of the possible $Y_c\bar{K}^{(*)}$ molecular states}

In this work, we further extend our study to the $\Lambda_c\bar{K}^{(*)}$ and $\Sigma_c\bar{K}^{(*)}$ systems, the corresponding OBE effective potentials can be related to those for the $\Lambda_c{K}^{(*)}$ and $\Sigma_c{K}^{(*)}$ systems by the $G-$parity rule \cite{Klempt:2002ap}, i.e.,
\begin{eqnarray}
V_{B_1\bar{M}_2\to B_3\bar{M}_4} &=& (-1)^{G_E}V_{B_1M_2\to B_3M_4},
\end{eqnarray}
where ${G_E}$ stands for the $G-$parity for the exchanged meson in the $B_1M_2\to B_3M_4$ process, notations $\bar{M}_i$ and ${M}_i$ correspond to the anti-mesons and mesons, respectively. Therefore, the effective potentials from the $\omega$ and $\pi$ exchanges are in completely contrast between the $Y_cK^{(*)}$ and the $Y_c\bar{K}^{(*)}$ systems.

\renewcommand\tabcolsep{0.2cm}
\renewcommand{\arraystretch}{1.6}
\begin{table*}[!hbtp]
\caption{The $\Lambda$ dependence of the obtained bound-state solutions (the binding energy $E$ and the root-mean-square radius $r_{RMS}$) for the single $Y_c\bar{K}^{(*)}$ systems. Here, $E$, $r_{RMS}$, and $\Lambda$ are in units of MeV, fm, and GeV, respectively. } \label{num4}
\begin{tabular}{cccc|cccc|cccc}
\toprule[1pt]\toprule[1pt]
  Systems    &$\Lambda$  &$E$   &$r_{RMS}$
 &Systems     &$\Lambda$  &$E$   &$r_{RMS}$ &Systems     &$\Lambda$  &$E$   &$r_{RMS}$\\\hline
 $\Lambda_c\bar{K}[1/2(1/2^-)]$   &\ldots  &\ldots  &\ldots
      &$\Sigma_c\bar{K}[1/2(1/2^-)]$   &1.30 &$-0.54$  &5.55
      &$\Sigma_c\bar{K}^*[3/2(1/2^-)]$   &4.40   &$-0.10$   &6.58\\
 $\Lambda_c\bar{K}^*[1/2(1/2^-)]$   &\ldots  &\ldots  &\ldots
      &     &1.35   &$-3.21$   &2.97
      &&4.60   &$-0.40$   &5.43\\
 $\Lambda_c\bar{K}^*[1/2(3/2^-)]$   &\ldots  &\ldots  &\ldots
      &&1.40    &$-7.92$   &1.99
      &&4.80   &$-0.84$   &4.34\\
 $\Sigma_c\bar{K}[3/2(1/2^-)]$   &\ldots &\ldots  &\ldots
      &&1.45   &$-14.56$   &1.52
      &&5.00   &$-1.47$   &3.47\\\hline
 $\Sigma_c\bar{K}^*[1/2(1/2^-)]$   &0.85   &$-0.37$   &5.42
 &$\Sigma_c\bar{K}^*[1/2(3/2^-)]$  &0.96   &$-0.40$   &5.64
 &$\Sigma_c\bar{K}^*[3/2(3/2^-)]$   &1.50   &$-0.75$   &4.49\\
    &0.90   &$-3.49$   &2.35       &&1.02    &$-3.26$   &2.66
     &&1.70     &$-4.46$    &2.11 \\
    &0.95   &$-9.56$   &1.54       &&1.08    &$-9.90$   &1.69
    &&1.90     &$-11.09$   &1.42\\
    &1.00   &$-18.11$   &1.20      &&1.14    &$-21.21$  &1.25
     &&2.10     &$-20.62$   &1.09\\
\bottomrule[1pt]
\bottomrule[1pt]
\end{tabular}
\end{table*}

In the following, we also perform the single channel analysis and the coupled channel analysis on the $Y_c\bar{K}^{(*)}$ systems. We summary the corresponding numerical results in Table \ref{num4} and Table \ref{num5}, respectively.

As shown in Table \ref{num4}, we collect the bound state properties for the single $Y_c\bar{K}^{(*)}$ systems. In the cutoff region $0.80\leq\Lambda\leq5.00$ GeV, we can obtain five loosely bound states, the $\Sigma_c\bar{K}$ bound state with $I(J^P)=1/2(1/2^-)$, the $\Sigma_c\bar{K}^*$ states with $I(J^P)=1/2(1/2^-)$, $1/2(3/2^-)$, $3/2(1/2^-)$ and $3/2(3/2^-)$. Among these five bound states, we cannot recommend the $\Sigma_c\bar{K}^*$ state with $3/2(1/2^-)$ as a good hadronic molecular candidate, as the cutoff value is too far away from the empirical value $\Lambda\sim1.00$ GeV. For the remaining four bound states, we conclude that they can be prime hadronic molecular candidates when we take the same cutoff criterion in the deuteron.

\renewcommand\tabcolsep{0.2cm}
\renewcommand{\arraystretch}{1.7}
\begin{table*}[!hbtp]
\caption{The bound state solutions (the binding energy $E$, the root-mean-square radius $r_{RMS}$, and the probabilities $P_i(\%)$ for all the discussed channels) for the coupled $Y_c\bar{K}^{(*)}$ systems. Here, $E$, $r_{RMS}$, and $\Lambda$ are in units of MeV, fm, and GeV, respectively. The dominant channels are labeled in a bold manner.}\label{num5}
\begin{tabular}{c|ccc|cccccc}
\toprule[1pt]
 $I(J^{P})$      &$\Lambda$    &$E$       &$r_{RMS}$
      &$\Lambda_c\bar K^*({}^2S_{1/2})$   &$\Lambda_c\bar K^*({}^{4}D_{1/2})$     &$\Sigma_c\bar K^*({}^{2}S_{1/2})$      &$\Sigma_c\bar K^*({}^{4}D_{1/2})$\\\midrule[1pt]
$1/2(1/2^-)$    &1.38   &$-0.90$   &2.98  &\bf{81.97}  &0.18   &16.99   &0.86\\
         &1.40   &$-4.77$   &1.21  &\bf{66.23}  &0.34    &31.86   &1.57\\
         &1.42   &$-9.42$   &0.88  &\bf{56.07}   &0.45    &\bf{41.46}   &2.02\\
         &1.44   &$-14.44$  &0.75  &\bf{49.04}   &0.53    &\bf{48.10}  &2.33\\
                \hline
$I(J^{P})$      &$\Lambda$    &$E$       &$r_{RMS}$
      &$\Lambda_c\bar K^*({}^4S_{3/2})$   &$\Lambda_c\bar K^*({}^{2}D_{3/2})$     &$\Lambda_c\bar K^*({}^{4}D_{3/2})$   &$\Sigma_c\bar K^*({}^{4}S_{3/2})$      &$\Sigma_c\bar K^*({}^{2}D_{3/2})$   &$\Sigma_c\bar K^*({}^{4}D_{3/2})$
      \\\hline
$1/2(3/2^-)$  &1.38  &$-2.71$  &1.51
                 &\bf{40.22}  &0.27  &2.16 &\bf{48.94}  &0.46 &7.96\\
       &1.39  &$-8.95$  &0.86 &28.77  &0.32  &2.53 &\bf{58.55}  &0.53 &9.29\\
       &1.40   &$-15.88$  &0.72 &23.23  &0.34  &2.69 &\bf{63.27}  &0.56 &9.90\\
       &1.41   &$-23.20$ &0.66  &19.83  &0.36  &2.77 &\bf{66.19}  &0.58 &10.27\\\hline
  $I(J^P)$ &$\Lambda$    &$E$  &$r_{RMS}$
      &$\Sigma_c\bar K({}^2S_{1/2})$    &$\Lambda_c\bar K^*({}^2S_{1/2})$   &$\Lambda_c\bar K^*({}^{4}D_{1/2})$     &$\Sigma_c\bar K^*({}^{2}S_{1/2})$      &$\Sigma_c\bar K^*({}^{4}D_{1/2})$\\\hline
 $1/2(1/2^-)$
 &0.87   &$-0.34$    &6.18    &\bf{98.57}   &0.90    &0.40      &0.04     &0.09\\
 &0.91   &$-3.71$    &2.85    &\bf{97.15}   &1.90    &0.68      &0.07     &0.19\\
 &0.95   &$-11.61$   &1.70    &\bf{95.52}   &3.15    &0.92      &0.09     &0.32\\
 &0.99   &$-25.39$   &1.21    &\bf{93.90}   &4.47    &1.09      &0.08   &0.46\\\hline
  $I(J^P)$ &$\Lambda$    &$E$  &$r_{RMS}$
      &$\Sigma_c\bar K({}^2S_{1/2})$    &$\Lambda_c\bar K^*({}^2S_{1/2})$   &$\Lambda_c\bar K^*({}^{4}D_{1/2})$     &$\Sigma_c\bar K^*({}^{2}S_{1/2})$      &$\Sigma_c\bar K^*({}^{4}D_{1/2})$\\\hline
$3/2(1/2^-)$
 &3.80   &$-0.54$    &5.47
        &\bf{99.43}      &\ldots      &\ldots     &0.28    &0.29\\
 &4.00   &$-2.67$    &3.12
        &\bf{99.00}      &\ldots      &\ldots     &0.50    &0.50\\
 &4.20   &$-6.91$    &2.00
        &\bf{98.42}      &\ldots      &\ldots     &0.82    &0.76\\
 &4.40   &$-14.02$   &1.43
        &\bf{97.68}      &\ldots      &\ldots     &1.23    &1.09\\
\bottomrule[1pt]
\end{tabular}
\end{table*}

When we consider the coupled channel analysis, we find four weakly bound states by varying the cutoff from 0.80 GeV to 5.00 GeV as shown in Table \ref{num5}. For the $\Lambda_c\bar{K}^*/\Sigma_c\bar{K}^*$ coupled bound state with $I(J^P)=1/2(1/2^-)$ and the $\Sigma_c\bar{K}/\Lambda_c\bar{K}^*/\Sigma_c\bar{K}^*$ coupled bound state with $I(J^P)=1/2(1/2^-)$, we can obtain the reasonable loosely bound state properties at the cutoff taken around 1.00 GeV, the dominant channels are the $\Lambda_c\bar{K}^*({}^2S_{1/2})$ and $\Sigma_c\bar{K}({}^2S_{1/2})$ channels, respectively. Thus, these two coupled bound states can be prime hadronic molecular candidates, which are mainly composed by the $\Lambda_c\bar{K}^*$ and $\Sigma_c\bar{K}$ states, respectively.

Compared to the bound states solutions for the single $\Lambda_c\bar{K}^*$ and $\Sigma_c\bar{K}$ systems with $1/2(1/2^-)$, we also find that the coupled channel effects play an important role in generating the $\Lambda_c\bar{K}^*$ state with $1/2(1/2^-)$. However, it contributes very little for the $\Sigma_c\bar{K}$ state with $1/2(1/2^-)$. Thus, the $\Sigma_c\bar{K}/\Lambda_c\bar{K}^*/\Sigma_c\bar{K}^*$ coupled bound state with $I(J^P)=1/2(1/2^-)$ predicted here is not a new bound state but has a close relation with the single $\Sigma_c\bar{K}$ molecule with $1/2(1/2^-)$.

For the $\Lambda_c\bar{K}^*/\Sigma_c\bar{K}^*$ coupled system with $I(J^P)=1/2(3/2^-)$, its dominant channel is the $\Sigma_c\bar{K}^*({}^4S_{3/2})$. As shown in Table \ref{num5}, its size is much smaller than those coupled channel bound states mainly made up by the lowest system. As the dominant channel is the $\Sigma_c\bar{K}^*({}^4S_{3/2})$, this bound state has a close relation to the $\Sigma_c\bar{K}^*$ molecule with $1/2(3/2^-)$.

For the $\Sigma_c\bar{K}/\Sigma_c\bar{K}^*$ coupled system with $I(J^P)=3/2(1/2^-)$, we can obtain the bound state solution as the cutoff reaches up to 3.80 GeV. Obviously, the cutoff applied here is deviated from the reasonable value 1.00 GeV. It cannot be a good molecular candidate.

All in all, our results can predict five $Y_c\bar{K}^{(*)}$ type hadronic molecular candidates, the coupled $\Lambda_c\bar{K}^*/\Sigma_c\bar K^*$ molecule with $1/2(1/2^-)$, the $\Sigma_c\bar{K}/\Lambda_c\bar{K}^*/\Sigma_c\bar K^*$ molecule with $1/2(1/2^-)$, the $\Sigma_c\bar{K}^*$ molecules with $1/2(1/2^-,3/2^-)$, and $3/2(3/2^-)$, where the coupled channel effects play a vital role in binding the coupled $\Lambda_c\bar{K}^*/\Sigma_c\bar{K}^*$ state with $1/2(1/2^-)$. Their important two-body strong decay channels are summarized as follows, i.e.,
\begin{eqnarray}
\Sigma_c\bar K[1/2(1/2^-)] &\to& \left\{\Lambda_c\bar K, \Xi_c^{(')}\pi\right\},\nonumber\\
\Lambda_c\bar{K}^*/\Sigma_c\bar{K}^*[1/2(1/2^-)] &\to&
  \left\{\Lambda_c\bar K, \Sigma_c\bar K, D\Lambda, D\Sigma, \Xi_c^{(')}\pi, \Xi_c^{(')}\eta\right\},\nonumber\\
\Sigma_c\bar{K}^*[1/2(1/2^-)] &\to&
   \left\{\Lambda_c\bar{K}^{(*)}, \Sigma_c\bar{K}, D^{(*)}\Lambda, D^{(*)}\Sigma, \right.\nonumber\\
   && \left. \Xi_c^{(')}\pi, \Xi_c^{(')}\eta, \Xi_c\rho, \Xi_c\omega\right\},\nonumber\\
\Sigma_c\bar{K}^*[1/2(3/2^-)] &\to&
   \left\{\Lambda_c\bar{K}^{*}, \Sigma_c^*\bar K, D^*\Lambda, D^*\Sigma,\right.\nonumber\\
   && \left. \Xi_c\rho, \Xi_c\omega, \Xi_c^*\pi, \Xi_c^*\eta\right\},\nonumber\\
\Sigma_c\bar{K}^*[3/2(3/2^-)] &\to& \left\{\Sigma_c^*\bar{K}, D^*\Sigma, \Xi_c\rho, \Xi_c^*\pi\right\}.\nonumber
\end{eqnarray}


\section{Summary}\label{sec4}

The study of the exotic states is an important and interesting issue in the hadron physics. Searching for the hadronic molecular states can not only enrich the family of the exotic states, but also help us to understand the essential hadron-hadron interactions. Very recently, the LHCb collaboration observed two open heavy flavor multiquarks $T_{c\bar s}^{a0(++)}$. Their near threshold behavior inspires the isovector $D^*K^*$ molecular explanations to them. In our former paper, we found the $D^*K^*$ state with $I(J^P)=1(0^+)$ can be possible molecular candidate by adopting the OBE effective potentials~\cite{Chen:2016ypj}.

\begin{figure}[!htbp]
\center
\includegraphics[width=3.4in]{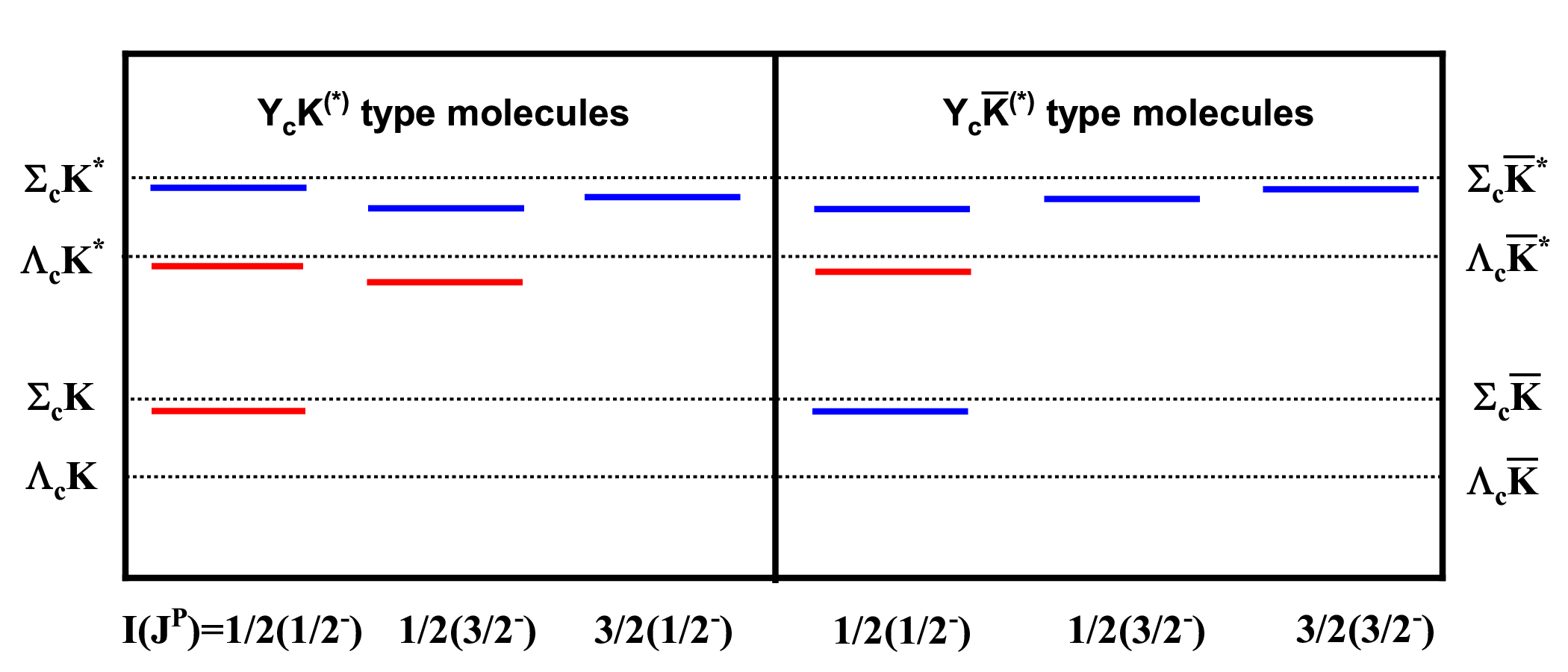}\\
\caption{A summary of the predicted $Y_cK^{(*)}$ and $Y_c\bar{K}^*$ molecular candidates. Here, the red and blue lines label the molecular candidates predicted by the single channel analysis and the coupled channel analysis. respectively.}\label{mass}
\end{figure}

In this work, we extend our study on the interactions between the $S-$wave charmed baryon $Y_c=(\Lambda_c,\Sigma_c)$ and the strange meson $K^{(*)}$ by using the OBE model, and we consider both of the $S-D$ wave mixing effects and the coupled channel effects. As shown in Figure \ref{mass}, our results indicate the single $\Sigma_cK^*$ states with $I(J^P)=1/2(1/2^-)$, $1/2(3/2^-)$ and $3/2(1/2^-)$ can be good open charm molecular candidates. When we further consider the coupled channel effects, we can predict another three prime open charm molecular candidates, i.e., the coupled $\Lambda_cK^*/\Sigma_cK^*$ molecular states with $1/2(1/2^-)$ and $1/2(3/2^-)$, and the coupled $\Sigma_cK/\Lambda_cK^*/\Sigma_cK^*$ molecular state with $1/2(1/2^-)$, where the dominant channels correspond to the $\Lambda_cK^*({}^2S_{1/2})$, $\Lambda_cK^*({}^4S_{3/2})$, and $\Sigma_cK({}^2S_{1/2})$, respectively. And the coupled channel effects play the essential role in binding these three coupled channel molecular candidates.

As a byproduct, we further study the $Y_c\bar{K}^{(*)}$ interactions in the same model. As shown in Figure \ref{mass}, we can predict the existences of the $Y_c\bar{K}^{(*)}$ type hadronic molecular states, i.e., the $\Sigma_c\bar{K}$ molecule with $I(J^P)=1/2(1/2^-)$, the $\Sigma_c\bar{K}^*$ molecules with $1/2(1/2^-)$, $1/2(3/2^-)$, and $3/2(3/2^-)$, the coupled $\Lambda_c\bar{K}^*/\Sigma_c\bar K^*$ molecule with $1/2(1/2^-)$, and the $\Sigma_c\bar{K}/\Lambda_c\bar{K}^*/\Sigma_c\bar K^*$ molecule with $1/2(1/2^-)$. We expect the experimentalists to search for these predicted open charm molecular pentaquarks.

\section*{ACKNOWLEDGMENTS}

R. C. is supported by the Xiaoxiang Scholars Programme of Hunan Normal University.

\end{document}